%% file: main.tex
\documentclass[12pt,letterpaper]{article}
\usepackage[utf8]{inputenc}
\usepackage[margin=1in]{geometry}
\usepackage[english]{babel}
\usepackage{geometry}
\usepackage{dcolumn}
\usepackage{float}
\usepackage{comment}
\usepackage{setspace}
\usepackage[dvipsnames]{xcolor}
\usepackage{natbib}
\usepackage{indentfirst}
\usepackage{amsfonts}
\usepackage{graphicx}
\usepackage{subcaption}
\usepackage{natbib}
\usepackage{caption}
\usepackage{tablefootnote} 
\usepackage{makecell}
\usepackage{amssymb}
\usepackage{latexsym}
\usepackage{amsmath, amsfonts,amssymb, amsthm, euscript,makeidx,color,mathrsfs}
\usepackage{enumerate}
\usepackage[bookmarksopen=true]{hyperref}
\usepackage{bookmark}
\usepackage{booktabs}
\usepackage{listings}
\usepackage{appendix}
\usepackage{array}
\newcommand{\PreserveBackslash}[1]{\let\temp=\\#1\let\\=\temp} 
\newcolumntype{C}[1]{>{\PreserveBackslash\centering}p{#1}} 
\newcolumntype{R}[1]{>{\PreserveBackslash\raggedleft}p{#1}} 
\newcolumntype{L}[1]{>{\PreserveBackslash\raggedright}p{#1}} 
\hypersetup{colorlinks=true,linkcolor=blue}
\usepackage{listings}
\definecolor{dgray}{gray}{0.35} 
\definecolor{lgray}{gray}{0.95} 

\title{Thinking inside the bounds: Improved error distributions for indifference point data analysis and simulation via beta regression using common discounting functions}
\author{Mingang Kim, Mikhail N. Koffarnus, Christopher T. Franck}
\date{}

\begin{document}
\doublespacing
\maketitle
\input{abstract.tex}
\input{Intro.tex}
\input{Methods.tex}

\input{Results.tex}
\input{Discussion.tex}

\input{Appendix.tex}

\bibliographystyle{abbrvnat}
\bibliography{reference}

\end{document}

%% file: abstract.tex
\section*{Abstract}

Standard nonlinear regression is commonly used when modeling indifference points due to its ability to closely follow observed data, resulting in a good model fit. However, standard nonlinear regression currently lacks a reasonable distribution-based framework for indifference points, which limits its ability to adequately describe the inherent variability in the data. Software commonly assumes data follow a normal distribution with constant variance. However, typical indifference points do not follow a normal distribution or exhibit constant variance. To address these limitations, this paper introduces a class of nonlinear beta regression models that offers excellent fit to discounting data and enhance{s} simulation-based approaches. Th{is} beta regression model can accommodate popular discounting functions. {This work proposes three specific advances. First, our model automatically captures non-constant variance as a function of delay. Second, our model improves simulation-based approaches since it obeys the natural boundaries of observable data, unlike the ordinary assumption of normal residuals and constant variance.} {Finally, we introduce a scale-location-truncation trick that allows beta regression to accommodate observed values of zero and one.} A comparison between beta regression and standard nonlinear regression reveals close agreement in the estimated discounting rate k obtained from both methods. 

%% file: Intro.tex
\section{Introduction}

Delay discounting refers to the phenomenon where the subjective value of a reward diminishes as the delay to its receipt increases. Individuals with high impulsivity tend to exhibit a stronger inclination towards immediate gratification, demonstrating higher rates of delay discounting. In contrast, individuals with low impulsivity demonstrate a greater ability to exercise self-control and opt for delayed, larger rewards. Therefore, delay discounting serves as an essential behavior analytic marker, offering insights into a range of problematic behaviors including drug dependence, gambling dependence, and obesity. More details about delay discounting related with problematic behaviors can be found in \citep{bickel2019excessive}. The purpose of this paper is to suggest a modeling approach that {more carefully} characterizes variability and obeys natural boundaries of delay discounting indifference points. 

 {An indifference point at a given delay describes the proportion of the larger later reward that an individual is willing to accept to receive reward sooner.} For example, if someone would be just as happy with \$87 today as \$100 in a week, the indifference point at a delay of one week is 0.87. {We focus on discounting tasks where the upper bound A, (i.e. the larger later amount, \$100 in this example) for indifference points is pre-set, so observed indifference points can be rescaled to be bounded bounded between 0 and 1 simply by dividing by A.}  We suggest a nonlinear beta regression approach that can be applied to common discounting functions. Further, we extend beta regression model by developing a Scale Location Truncated {(SLT)} beta regression {model} to accommodate 0 and 1 values, which cannot easily be incorporated with typical beta regression approaches.

\begin{figure}[H]
  \centering
  \begin{subfigure}[b]{0.44\textwidth}
    \centering
    \includegraphics[width=\textwidth]{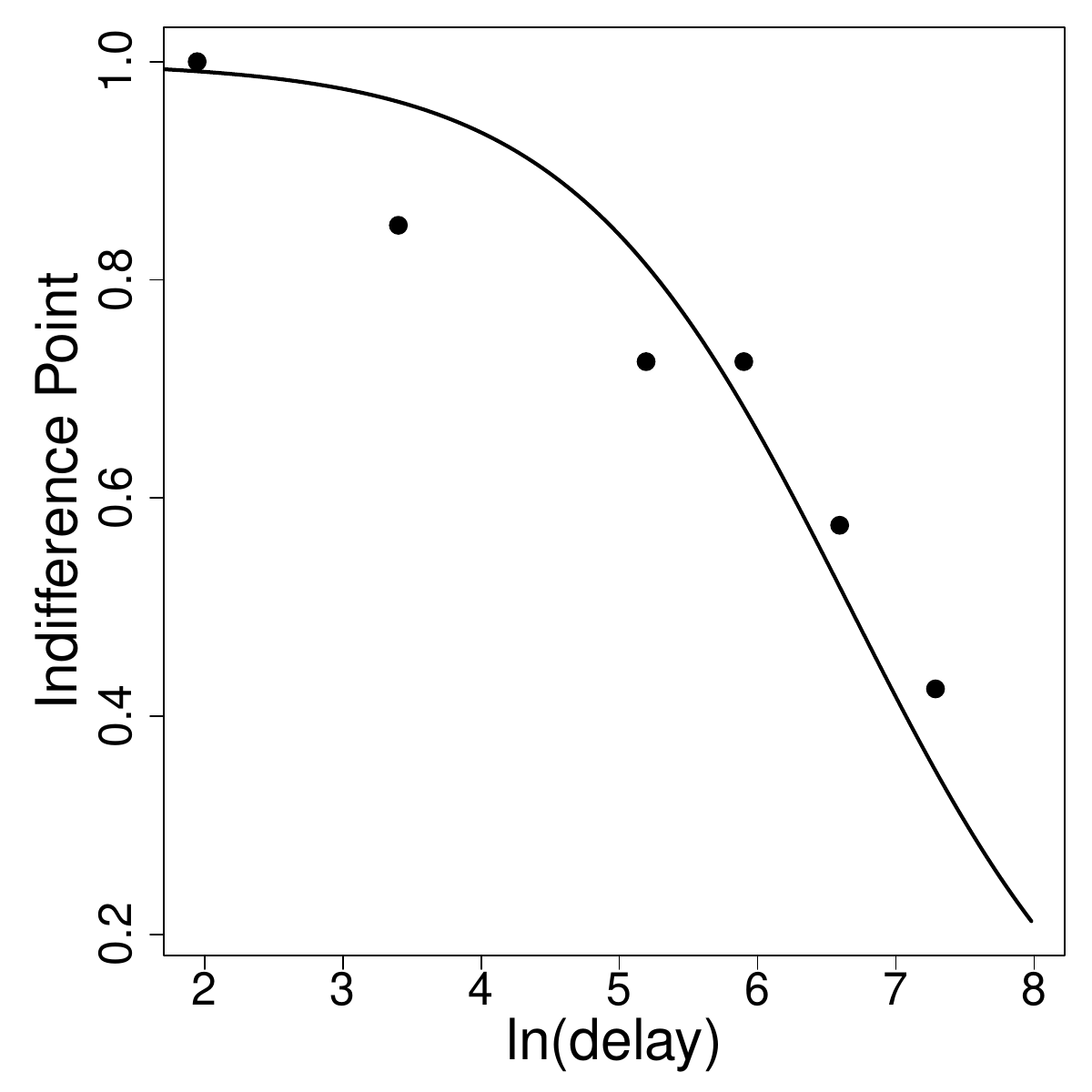}
  \end{subfigure}
  \hfill
  \begin{subfigure}[b]{0.52\textwidth}
    \centering
    \includegraphics[width=\textwidth]{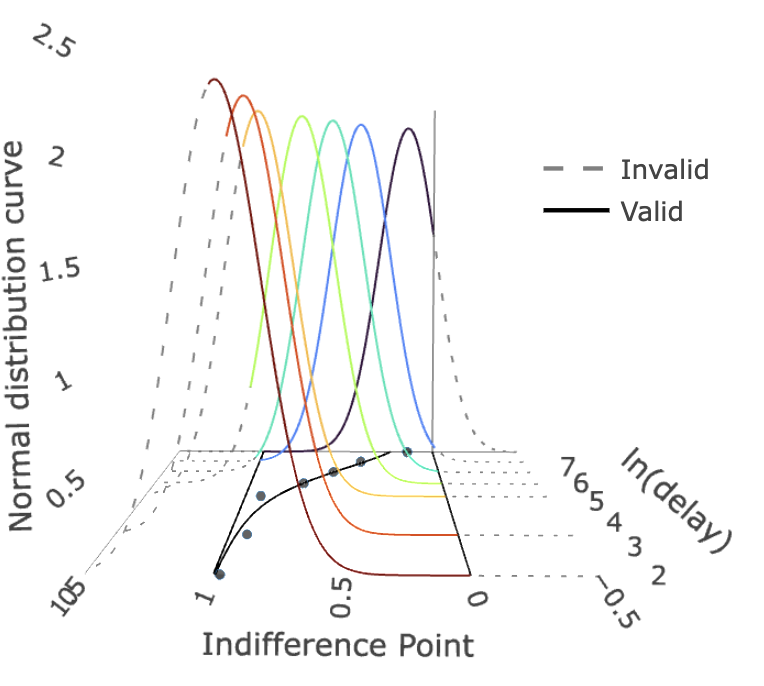}
  \end{subfigure}
  \caption{Indifference points \textcolor{black}{for participant} 4 \citep{jarvis2019effects} and \textcolor{black}{NLS hyperbolic discounting function} fit to the data (left panel), and normal distribution curve at each delay \textcolor{black}{for participant} 56 (right panel). Variance in normal curves is estimated on the basis of this model fit. The right panel of the plot shows the the problem \textcolor{black}{with} using \textcolor{black}{the} normal distribution \textcolor{black}{to describe variance} for indifference \textcolor{black}{point} data. The normal distribution includes probability mass for data outside of 0 and 1\textcolor{black}{,} which is invalid. The dotted lines in the right panel indicate invalid values.}
  \label{fig:fig1}
\end{figure}

The {hyperbolic discounting function \citep{mazur1987adjusting}} is a widely used model for {characterizing choice behavior in delay discounting studies}. It can be expressed as 

\begin{equation}\label{eq:eq1}
    E(y) = \frac{1}{1+k\cdot D} 
\end{equation}

where $D$ is {delay}, $k$ is the unknown {discounting rate} parameter to be estimated {using} data, $y$ is indifference point, and $E(y)$ indicates {the expected value} of $Y$, i.e, the value of the regression {line} for $k$ and $D$. The left panel of Figure \ref{fig:fig1} shows an example of a series of indifference points with the {hyperbolic discounting function} plotted. The nonlinear least squares (NLS) method  \citep{bates1988nonlinear} is commonly used to fit the {hyperbolic discounting function}. NLS estimates value of $k$ as the value that is the closest to the observed data {by minimizing the sum of squared residuals.} However, NLS does not assume a specific probability distribution {for observed data, and thus alone NLS does not prescribe a specific way to characterize variability in the data.} In statistics, capturing variability inherent in the observed data is important because it enables accurate estimation and prediction. In addition, while a common assumption in regression modeling is that residual error follows a normal distribution, this assumption does not hold for delay discounting. Data simulation based on real world delay discounting data \citep{jarvis2019effects} shows this clearly. When simulating data with normal residuals, 19.28\% of the indifference points were invalid, and in 76.3\% of cases, at least one subject had an invalid point when conducting 1000 Monte Carlo simulations. By contrast, simulating data from the beta distribution did not result in any invalid points. More detail of this Simulation can be found in Section 2.4 and Figure \ref{fig:fig7} in Section 3. 

Figure \ref{fig:fig1} show{s} the {hyperbolic discounting function} fit for subjects 4 and 56 from \citep{jarvis2019effects}. The left {panel} of Figure \ref{fig:fig1} is {the} {hyperbolic discounting function} fit {for} subject 4 . We can see that the indifference points {range} from [0, 1]. In this paper, [a, b] indicates a bounded range between a and b, including a and b. The right panel of Figure \ref{fig:fig1} shows {the} normal distribution curve {for each} delay. The normal distribution {curves are centered at the regression line,} and variance is estimated from the {model fit, i.e., how close the observed indifference points are to the regression line}. From the plot, we can see that the normal curve{s are} not bounded {and they stretch past the [0,1] boundaries}. {This indicates that} the {actual} distribution of residual error {for delay discounting analysis is not well characterized by the} normal distribution. 

By contrast, the beta regression model \citep{ferrari2004beta} is based on the beta distribution, which describes data on a range between 0 and 1. This coincides with the range of the indifference points. The beta distribution has two shape parameters which make the distribution flexible. Like the normal curve, the beta distribution has a curve that express{es} probabilities in terms of area under the curve. In general, the normal curve and the beta curve (Figure \ref{fig:fig2}) are {examples of} probability density functions (PDF). Thus, {the PDF} of the beta distribution describe{s} the behavior of the data, allowing users to compute the probabilities of various outcomes. In this case, indifference points are the outcomes. This means that beta regression can be widely used in various fields where response variable is continuous and in bounded between 0 and 1.  

There are some existing implementations of beta regression. The R package \textbf{betareg} \citep{cribari2010beta} fits {the linear} beta regression model. The limitations of this method are that {(i)} the response variable can not include 0 or 1 while those are possible values in some data sets {(including in the left panel of Figure \ref{fig:fig1})}. In addition, {(ii)} this package only fits linear relationships and thus it is not applicable to common nonlinear discounting functions. Another approach for beta regression to model nonlinear functions is in \citep{de2021beta}. This approach also does not accommodate 0 or 1 for the response variable.

Some transformation methods have been suggested {for cases where beta regression is desired but the observed data contains values of 0 or 1.} One such method is to transform data using $y'=(y-a)/(b-a)$ when the range is [a,b] and then transform $y''=[y'(N-1)+0.5]/N$ where N is {the} sample size \citep{smithson2006better}. However, this is not a model{-}based approach, and {the choice of transformation for zeroes in particular can be influential since the log function ascends so rapidly close to 0}. \cite{liu2015zoib} suggested zero/one inflated beta regression (ZOIB). The ZOIB consists of three segmented beta regression models that address response values at 0, 1, and values between 0 and 1. This mixture model-based approach may not be effective when there are a limited number of point mass values at 0 and 1 as we might expect in a typical discounting analysis. This is because mixture model requires relatively large number of observations. If sample size is not big enough {for both the endpoints and the interior of the space}, it is hard to reliably estimate the parameters.

\cite{berk2021appropriate} recognized the same problem with {non-normal} indifference points but they use a different approach from ours. {While we use {hyperbolic discounting function} which is nonlinear form, they transform the {hyperbolic discounting function} to a linear form. Thus, their response variable encompasses values ranging from zero upwards, with no upper bound which is different from our response range.} 

In this paper, we propose the beta distribution as a probability model for indifference points, which correctly adheres to the {[0,1] range for} delay discounting data. The de-identified data and code used for the analyses in this paper can be found as online supplemental materials accompanying this paper and at the following link:  \url{https://bitbucket.org/paper-code-kmg/slt_beta/src/main/}.

%% file: Methods.tex
\section{Methods}

{This section covers the} beta distribution, beta nonlinear regression, and SLT beta regression.  First we explore {the} beta density function, reparameterization, and log-likelihood function, and the reason {that standard} beta regression cannot accommodate 0 and 1 values. We will also develop a better understanding about building nonlinear beta regression models. We develop SLT beta regression and show how it can {accomodate indifference points with values of} 0 and 1. While other common discounting  functions {can} be used, we focus on beta nonlinear regression using the {hyperbolic discounting function}. {Finally}, we {describe} simulation methods based on real world data \citep{jarvis2019effects}. 

The data we analyze is from \cite{jarvis2019effects}. The data include 146 {participants} with 34 {individuals} that have at least one 0 or 1 indifference point. Participants are a subset of participants analyzed in the original study. The collection of discounting data in that study was an add-on procedure that started after the main study was already ongoing, so not {all participants} completed the task, and those who did complete the task did not necessarily do so at all time points. Among the 146 {participants}, we used 126 subjects who passed Johnson-Bickel Criteria \citep{johnson2008algorithm} to identify systematic delay discounters. 

\subsection{Beta distribution}

The beta curve, i.e, PDF is typically expressed as

\begin{equation}\label{beta_density}
     f(y;\alpha, \beta)= \frac{\Gamma(\alpha+\beta)}{\Gamma(\alpha)\Gamma(\beta)}y^{\alpha-1}(1-y)^{\beta-1}, \ \  0< y < 1
\end{equation}

where shape parameters $\alpha, \beta$ $>$ 0 and $\Gamma(\cdot)$ is the gamma function. 

\begin{figure}[ht]
    \centering
    \includegraphics[width=1\textwidth]{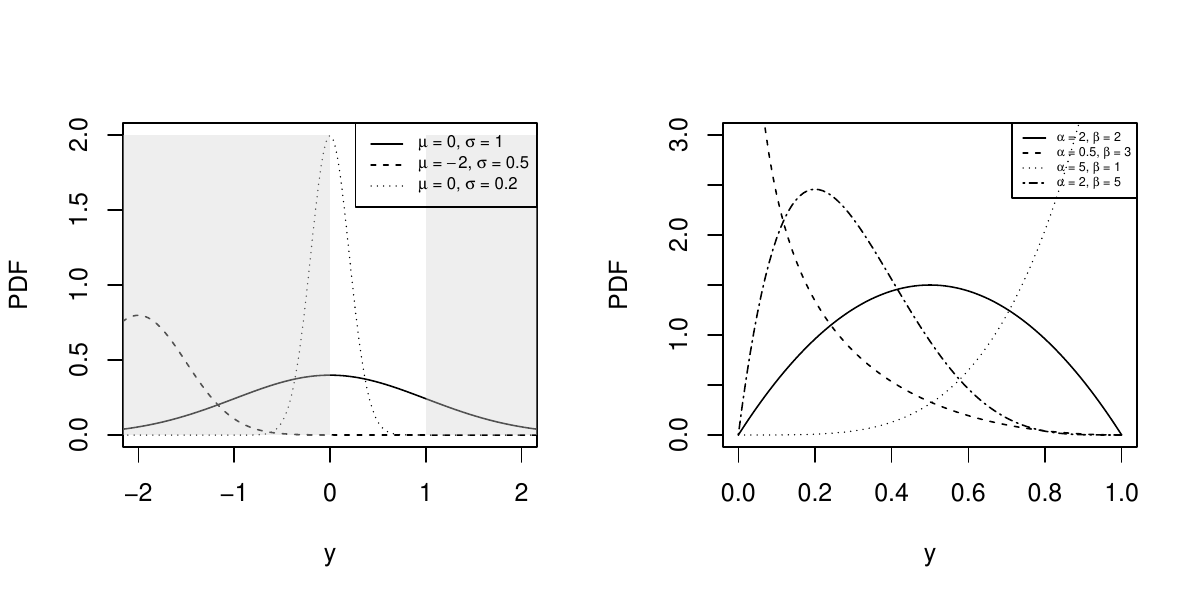}
    \caption{PDF plot of normal distribution (left) and PDF of beta distribution (right). {In the left panel of the plot,} the white region represents the range of indifference points, spanning values from 0 to 1, while the grey area represents the values outside of [0, 1]. {The right panel shows that} the beta distribution's PDF exhibits a variety of shapes contingent upon the values of its shape parameters, $\alpha$ and $\beta$, {and is constrained to be between zero and one}. }
    \label{fig:fig2}
\end{figure}

Figure \ref{fig:fig2} illustrates examples of normal and beta PDFs. The normal PDF exhibits different patterns based on its mean parameter $\mu$ and standard deviation $\sigma$. The beta PDF exhibits different patterns based on the values of the shape parameters $\alpha$ and $\beta$. However, in all scenarios, the {beta} density {equals 0 when the range is less than or equal to}  0 {or greater than or equal to } 1. On the other hand, PDF of normal distribution {is} bell-shaped, {can vary in both mean} and dispersion, and has positive values even beyond 0 and 1. 

\cite{ferrari2004beta} proposed a different {parameterization for the beta density that better accomodates regression problems}. This reparameterization rewrites {the beta density in terms} of mean {parameter} {$\mu$ and} scale {parameter} $\phi$. {The mean $\mu$ is set equal to the value of the hyperbolic discounting function (or other discounting function preferForestGreen by the user}.  {The scale parameter} $\phi$ accommodates non-constant variance {in indifference points as a function of delay}. {For the ordinary normal case, variance of y equals $\sigma^2$ regardless of delay and beta regression.  However, in the beta distribution, variance of y equals to $\mu(1-\mu)/(1+\phi)$, which varies depending on delay.}
 
\subsection{Scale Location Truncated beta regression (SLT beta regression)}

By default, the beta distribution cannot accommodate values at 0 or 1. {However, values of 0 and 1 may appear among indifference points. While many common adjusting procedures do not produce 0 or 1 values, there are several cases in which these extreme values can show up in the data, and in some cases the experimental question may require us to establish whether indifference points are at 0 or 1, not just “close” to 0 or 1.} In order to give the beta density mass on 0 and 1 endpoint values, we propose a scale-location truncation (SLT) strategy. Truncating a density is typically used for restricting values to a specific range. An interesting property of {the} truncated beta density is that {it takes} positive values on their endpoints unlike the usual beta density. By using truncation on the beta density, we {have developed the SLT beta} density function which has mass on 0 and 1, unlike {the usual} beta density function. {This is shown in the right panel of Figure \ref{fig:fig8} in the Appendix.} Thus, our {SLT beta density function can be used to form a likelihood function, which can in turn be maximized to furnish estimates of discounting rate and variance, even when there are 0 and/or 1 values among the indifference points. The technical details of the SLT beta density function can be found in the} Appendix.  

\subsection{Nonlinear Beta Regression}

We {next pair} the {hyperbolic discounting function} \citep{mazur1987adjusting} {with} our {SLT} beta regression approach. {Recall that} Equation (\ref{eq:eq1}) is the {hyperbolic discounting function}. {To complete the specification of our SLT Beta regression model}, consider { the response variable $y_{ij}$ as the indifference point for $ith$ subject and $jth$ delay.} Let $y_{ij}$ be a random sample which follows $y_{ij} \sim \text{{SLT Beta}}(\mu_{ij}, \phi_i)$. {The mean of the distribution $\mu_{ij}$ depends on both the subject and the delay, and follows the hyperbolic discounting function. Thus,}

\begin{equation*}
    {\mu_{ij}=} \frac{1}{1+k_i \cdot D_j},
\end{equation*}

{where the $D_j$ variable is $jth$ delay, and we estimate discounting rate $k_i$ for participant $i$} using observed indifference point data. 
Instead of {$k_i$}, we {typically directly} estimate $\psi_i=ln(k_i)$, which implies $\mu_{ij}=\frac{1}{1+exp(\psi_i) \cdot D_j}$ and {$y_{ij} \sim B\left(\frac{1}{1+exp(\psi_i)\cdot D_j}, \phi_i \right )$.}

As previously mentioned, the {PDF and hence} log-likelihood of the beta {distribution} cannot accommodate 0 and 1. Thus, we use {our SLT} beta regression {approach} for estimating $\psi$ and $\phi$. {This is the log-likelihood for the SLT Beta regression model:} 

{
\begin{equation*}
    \mathcal{L}(y_{ij}; \psi_i, \phi_i, s, l)= \sum_{j=1}^{J} \ \  ln(\Gamma(\phi_i))-ln\left(\Gamma\left(\frac{\phi_i}{1+exp(\psi_i)\cdot D_j}\right )\right)-\ln\left(\Gamma \left(\left(1-\frac{1}{1+\exp(\psi_i)\cdot D_j}\right)\cdot  \phi_i \right)
 \right)
\end{equation*}
}
{
\begin{equation*}
    -\left(\frac{\phi_i}{1+exp(\psi_i)\cdot D_j}-1\right)\cdot ln\left(\frac{y_{ij}}{s}+l\right)+
    \left(\left(1-\frac{1}{1+exp(\psi_i)\cdot D_j}\right)\phi_i -1 \right)\cdot ln\left(1-\left(\frac{y_{ij}}{s}+l\right)\right),
\end{equation*}
}

{
\begin{equation*}
   -ln\left(F_y\left(\frac{1}{s}+l \right)-F_y(l)\right),
\end{equation*}
}

{and details of its derivation can be found in the Appendix.}

{In Equation (\ref{eq:eq8}), we introduce a reparameterization of the response variable, denoted as $g$. The variable $y_{ij}$ represents the original form of the response variable, while $g$ serves as a reparameterized version used for mathematical clarity. This reparameterization helps differentiate the response variable from other variables, such as $z$, introduced in Equation (\ref{eq:eq5}). Despite being represented differently, $g$ and $y_{ij}$ fundamentally refer to the same response variable.}

By maximizing $ \mathcal{L}(y_{ij}, \psi_i, \phi_i)$, we can estimate $\psi_i$ and  $\phi_i$ {for each participant.} In other words, using maximum likelihood estimation, we select values of $\psi_i$ and $\phi_i$ that make observed data most likely to have occurForestGreen {for each individual series of data.} {An area of future research would be to extend the SLT beta regression approach for hierarchical modeling.}

\subsection{Reanalysis of human discounting data}

{We use human discounting data \citep{jarvis2019effects} to assess the utility of the proposed SLT Beta regression approach with the following strategy. First, we conduct similar analyses as shown in Figure \ref{fig:fig1}, showing the SLT beta regression model fit and density curves within the space of indifference points to establish adequare model fit and distribution that obeys the [0,1] bounds. We then compare estimates of variability between NLS and SLT beta regression approaches to demonstrate the ability of SLT beta regression to model non-constant variance. We show that the empirical pattern of variability in indifference points as a function of delay matches the pattern shown using SLT beta regression. We show that estimates of $ln(k)$ are highly correlated between NLS and SLT beta regression, indicating high agreement the estimation of discounting rate between the methods. We plot model fits alongside data for several participants to show similarity in the regression lines produced by NLS and SLT Beta regression. Finally, we store estimates of $ln(k)$ and variance for use in our simulation study.}

\subsection{Simulation}

We conducted a simulation based on real world data \citep{jarvis2019effects}. Following the conventions of the Statistics literature, we denote the estimator of $k$ with $\hat{k}$. The simulation's algorithm is structuForestGreen as follows.\\ 

\textbf{Simulation from Normal distribution}

\begin{enumerate}
    \item Fit each subjects data to the {hyperbolic discounting function} using NLS, the current standard approach. {Store estimated values of $ln(\hat{k})$ and variance $\hat{\sigma}^2$.}
    \item Based on the estimated $\hat{k}$ from NLS, we simulate the $jth$ subject's indifference points from {a} normal distribution with parameters $\mu_{ij}$ and $\sigma_j^2$ {that are set equal to empirical estimates from the \citep{jarvis2019effects} data}. In other words, simulated data reflects {observed} regression lines and variance in data, but residuals follow {a} normal distribution {with constant variance, i.e., the typical assumptions for NLS modeling. Thus:} 

    \begin{equation}
        \mu_{ij}=\frac{1}{1+\hat{k}_{nls, {i}}*D_{{j}}}, \ \ \ \ \sigma_j^2=\sum_{i=1}^{d}\left(y_{ij}-\frac{1}{1+\hat{k}_{nls, {i}}*D_{{j}}}\right)^2/(d-1),
        \label{eq:eq2}
    \end{equation}

    where $D_{{j}}$ stands for the  ${j}th$ delay, $d$ is the number of delays,. and $\hat{k}_{nls, {i}}$ is the estimated $k$ from NLS of the ${i}th$ subject.
\end{enumerate}

\textbf{Simulation from Beta distribution}

\begin{enumerate}
    \item Fit each {participants'} data to the hyperbolic discounting function using {SLT} beta regression.
    \item Based on the estimated $\hat{k}_i$ and $\hat{\phi}_i$ from beta regression,  we simulate {the} $ith$ subject's indifference points from {the} beta distribution with parameters $\alpha_{ij}$ and $\beta_{ij}$ {as shown in Equation (\ref{beta_density}}). Thus,

    \begin{equation}
        \mu_{ij}=\frac{1}{1+\hat{k}_{beta, {i}}*D_{{j}}}, \ \ \ \ 
       \alpha_{ij}= \mu_{ij}\phi_{{i}} \ \ \ \ \beta_{ij}=(1-\mu_{ij})*\phi_{{i}},
       \label{eq:eq3}
    \end{equation}

 where $D_{{j}}$ stands for ${j}th$ delay point and $d$ is the number of delay points. $\hat{k}_{beta, {i}}$ is estimate $k$ from NLS of {i}th subject and $\phi_{{i}}$ is {the} estimated $\phi$ of {i}th subject.

\end{enumerate}

\subsubsection{Monte Carlo simulation}

Monte Carlo simulation employs repeated sampling to assess the properties of specific phenomena. In this paper, we use Monte Carlo simulation to learn the percentage of invalid indifferent points {outside of the [0,1] bounds generated by the normal distribution and beta distribution strategies described above}. We use 1000 as a replication number. {While it may be obvious to some readers that the SLT beta regression approach cannot generate data outside [0, 1], we include a Monte Carlo study to empirically validate that our SLT beta regression model correctly obeys the [0, 1] bounds.} {We demonstrate that the SLT beta regression approach does not produce invalid out-of-bounds data, and that simulating normal residuals does produce invalid data. Crucially, this indicates that SLT beta regression model more closely reflects empirical data and thus is the more suitable method when using Monte Carlo simulation-based approaches for the study of delay discounting.}

The process of Monte Carlo simulation is as follows. 
\begin{enumerate}
    \item Simulate the data as described above from normal and SLT beta distributions. Replicate the data generation 1000 times using the inputs from the \cite{jarvis2019effects} data.
    \item Examine the simulated indifference points for each subject to determine whether the values exceed 1 or fall below 0. Calculate the percentage of simulated data points exceeding 1 and those falling below 0 for each subject and at various delay points. Summarize overall percentage of invalid points and percentage of invalid points by delay. Compute the percentage of participants with at least one invalid indifference point. 
    
\end{enumerate}

%% file: Results.tex
\section{Results}

\begin{figure}[H]
  \centering
  \begin{subfigure}[b]{0.48\textwidth}
    \centering
    \includegraphics[width=\textwidth]{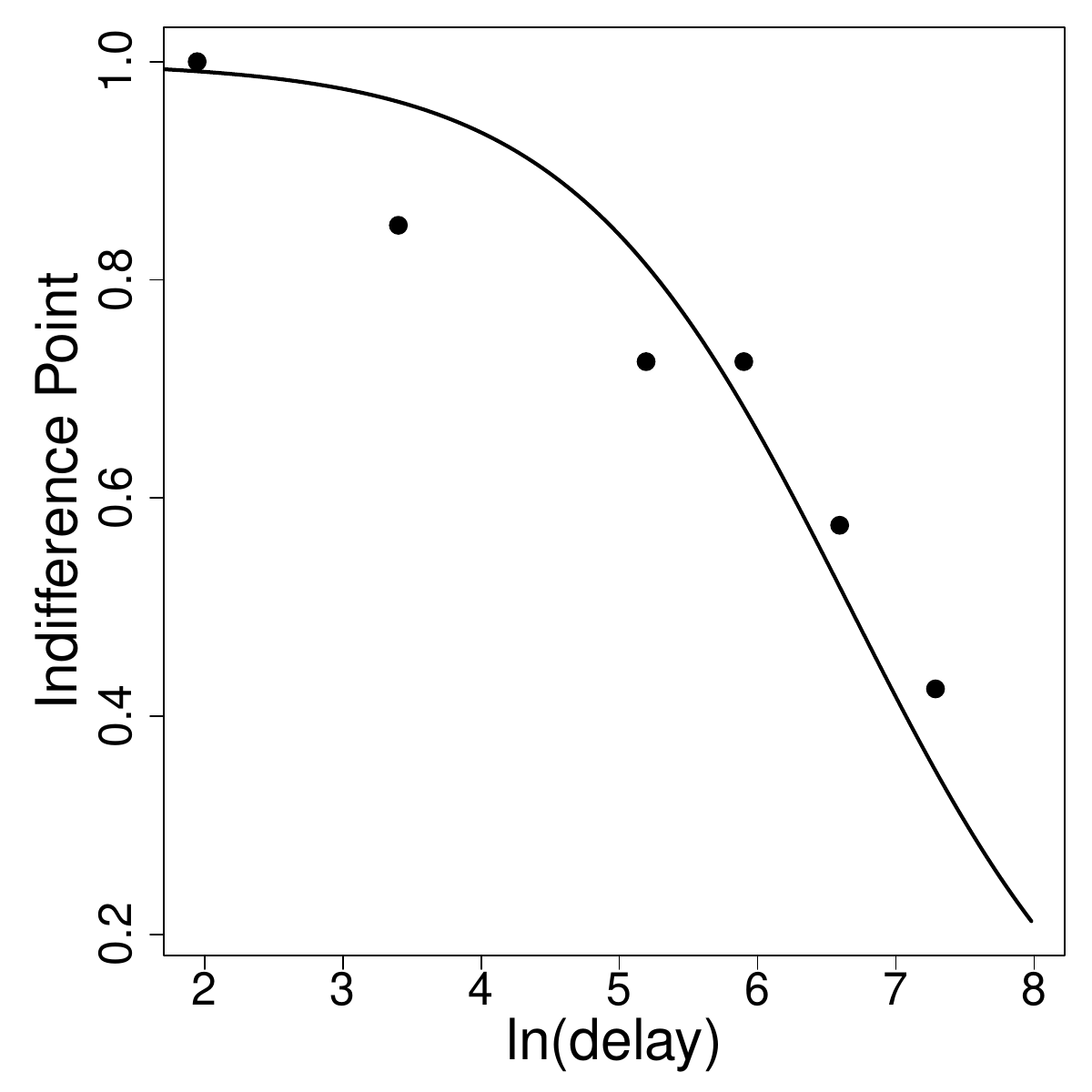}
  \end{subfigure}
  \hfill
  \begin{subfigure}[b]{0.50\textwidth}
    \centering
    \includegraphics[width=\textwidth]{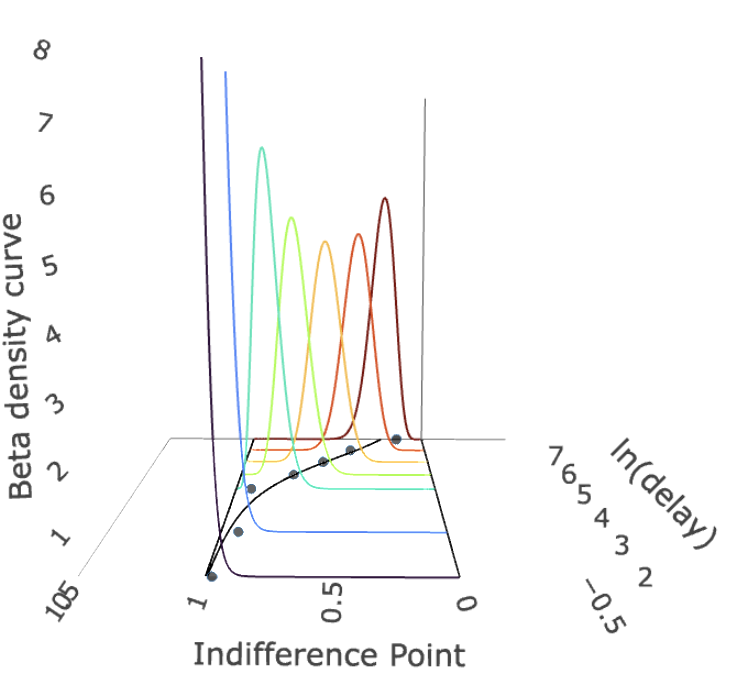}
  \end{subfigure}
  \caption{Indifference points and {SLT beta regression} fit to the data from {participant} 4  (left panel), and the PDF of the {SLT} beta distribution at each delay for subject 56 (right panel). In the right plot, unlike normal distribution case (Figure \ref{fig:fig1}) , {the SLT}  beta distribution {has positive probability between  0 and 1.}}
  \label{fig:fig3}
\end{figure}

The left panel of Figure \ref{fig:fig3} {shows the} model fit of {participant 4 from} \cite{jarvis2019effects}. The right panel of Figure \ref{fig:fig3}  shows how the {SLT} beta density {is} properly constrained in 0 and 1 when modeling indifference points {with the} {hyperbolic discounting function}. By contrast, {the} normal PDF in Figure \ref{fig:fig1}  gives positive probability for {values} over 1 and below 0.  

\begin{figure}[H]
    \centering
    \includegraphics[width=1\textwidth]{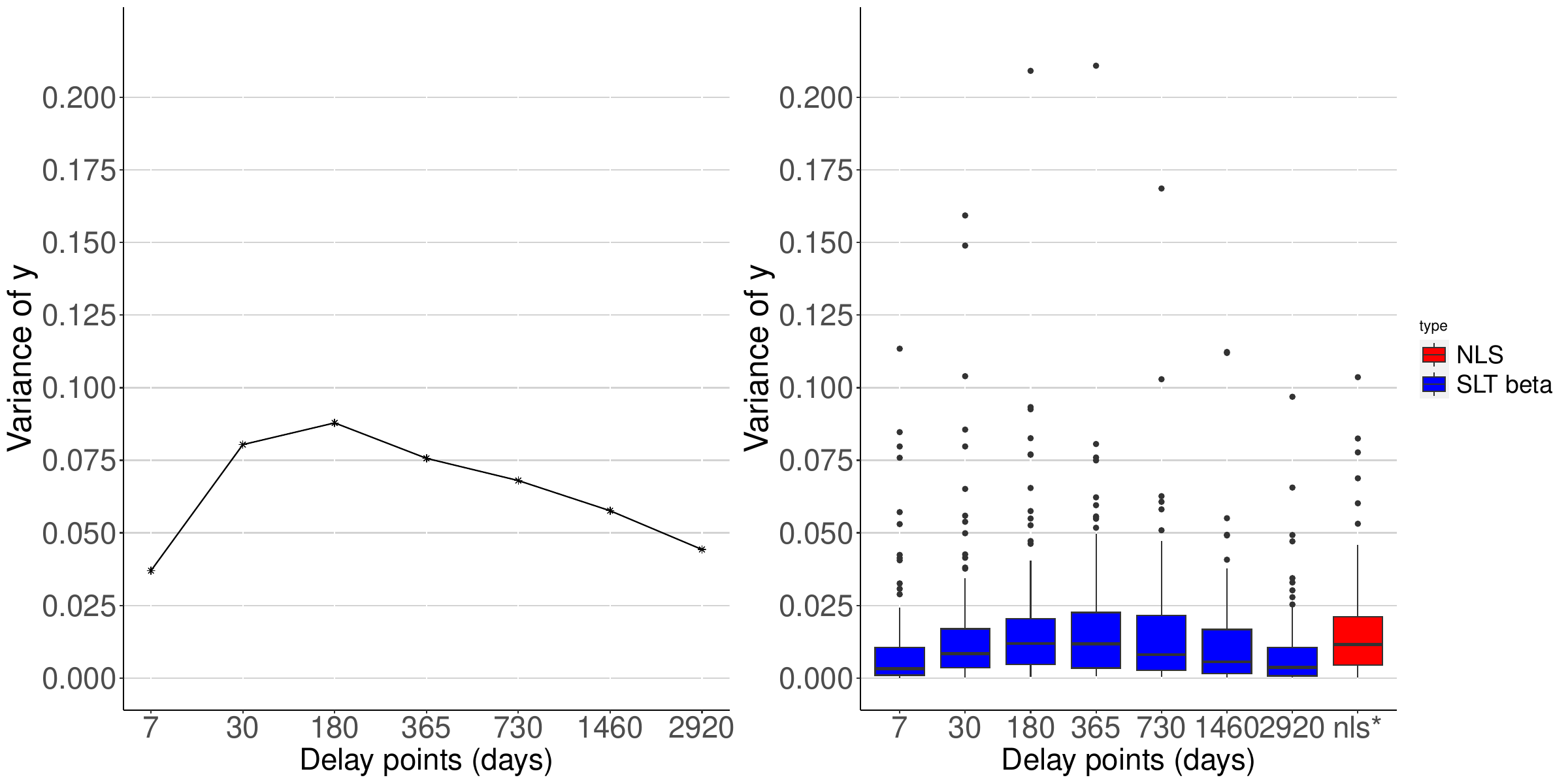}
    \caption{Variance among subjects at each delay (left panel). Box plot of model estimated variance from SLT beta regression at each delay and variance from NLS (right panel). Delays are graphed as equally spaced for ease of visualization.}
    \label{fig:fig4}
\end{figure}

The left panel of Figure \ref{fig:fig4}  shows variance among subjects at each delay. {This plot shows that} at low and high delay{s}, variance is {lower}, and {in the middle of the delay range}, variance is {higher}. {This is because low delays tend to elicit little discounting and there is an upper bound on indifference points at 1. Similarly, for very large delays, individuals tend to have indifference points near the boundary of 0 where there is little ``wiggle room'' and thus lower variance. By contrast, indifference points in the middle of the range are free to vary without bumping into any boundaries which is why empirically we tend to see larger variances here.} The right panel of Figure \ref{fig:fig4}  {shows the estimated variance} at each delay {according to the SLT beta regression} fit. On the right plot, while model estimated variance of SLT beta at each delay shows similar pattern with variance of subjects at fixed delay, model estimated variance from NLS has common variance at all delays. {Importantly, the NLS with constant variance approach tends to produce an estimate of variance that is larger than most of the delay-specific variances that can be accommodated by the SLT beta regression approach. NLS misses the pattern of variability inherent in the data.}

\begin{figure}[H]
    \centering
    \includegraphics[width=0.55\textwidth]{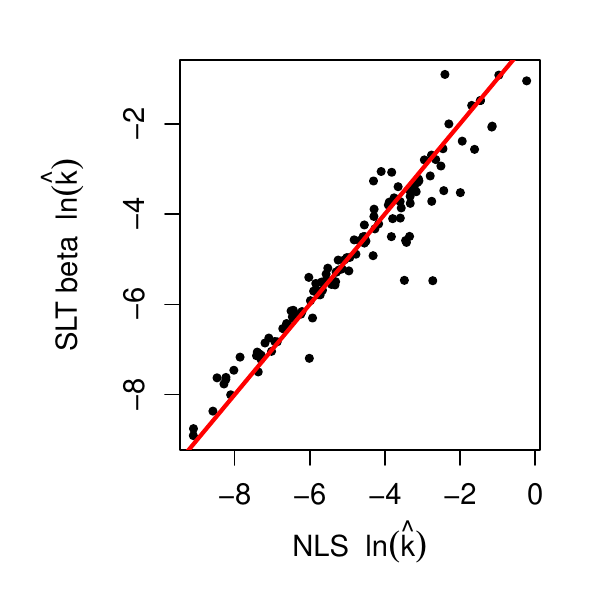}
    \caption{Scatter plot of ln($\hat{k}$). Horizontal axis is ln($\hat{k}$) from NLS and vertical axis is ln($\hat{k}$) from SLT {beta regression}. Red line indicates y=x.}
    \label{fig:fig5}
\end{figure}

In Figure \ref{fig:fig5}, we can see that points are closely located around the the line of equality (y=x). This demonstrates that the estimated values of {$ln(\hat{k})$} obtained from NLS and SLT beta regression are remarkably similar. Table \ref{tab:tabl} demonstrates the similarity between the distributions of $ln(k)$ obtained from NLS and SLT beta regression.

\begin{table}[H]
\centering
\begin{tabular}{crrrrrrr}
  \hline
  method & Min & Q1 & Median & Q3 & Max & Mean & SD \\ 
  \hline
NLS ln($\hat{k}$) & -9.10 & -6.32 & -4.88 & -3.37 & -0.23 & -4.86 & 1.94 \\ 
  SLT ln($\hat{k}$) & -8.91 & -6.23 & -4.99 & -3.62 & -0.91 & -4.92 & 1.76 \\ 
   \hline
\end{tabular}
   \caption{Summary statistics of ln(k) from NLS and SLT beta {regression} for subjects which satisfy the {Johnson-Bickel criteria \citep{johnson2008algorithm}}. Q1 indicates 25 percentile and Q3 indicate 75 percentile.}
   \label{tab:tabl}
\end{table}

\begin{figure}[H]
    \centering
    \includegraphics[width=1\textwidth]{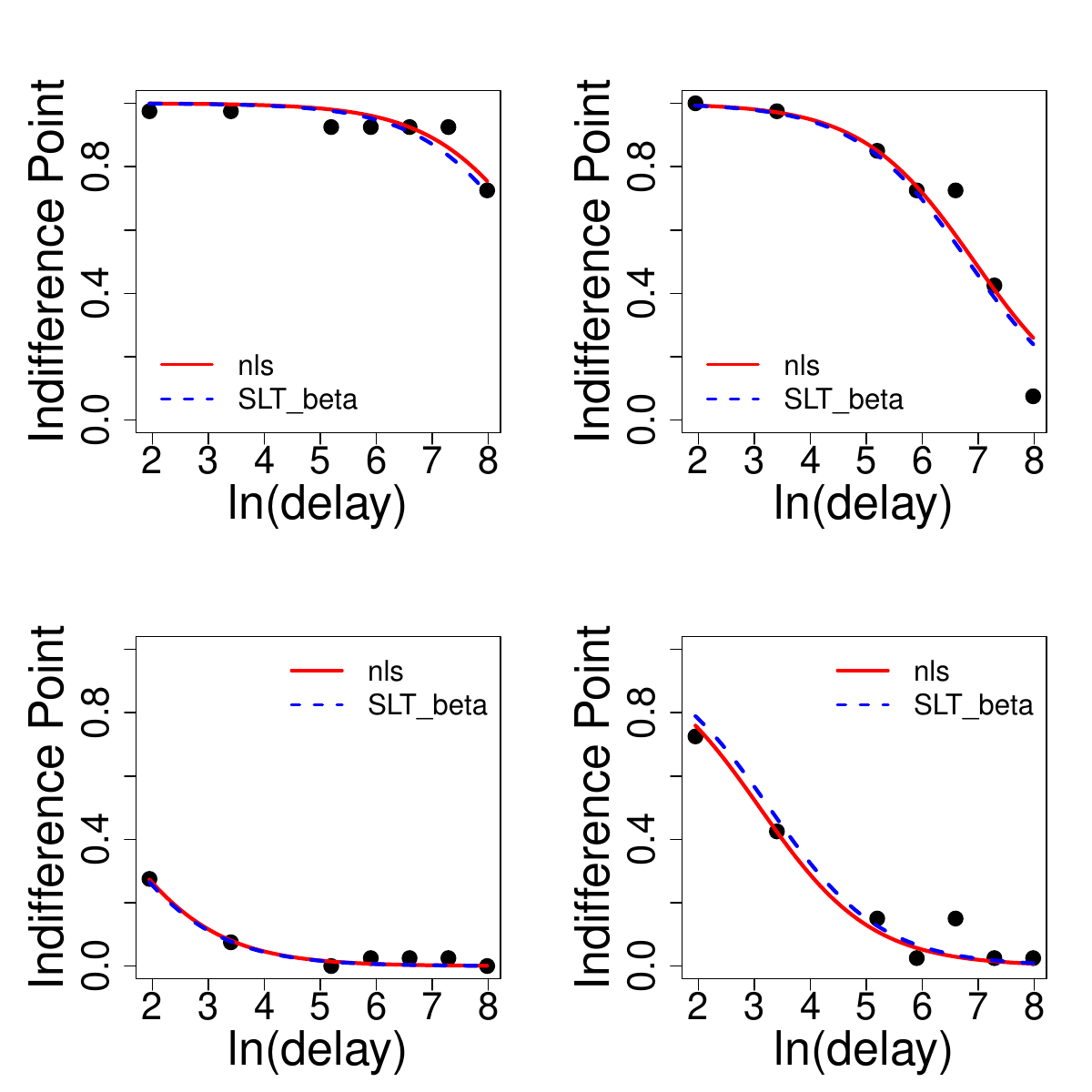}
    \caption{Model fits for subject 24 and 33 (top row), for subject 43, 7 (bottom row) using NLS and SLT beta regression.}
    \label{fig:fig6}
\end{figure}

Figure \ref{fig:fig6} shows model fits from NLS and SLT beta regression for indicated subjects. The four plots show that their fits are very close various in different settings.

\begin{figure}[H]
    \centering
    \includegraphics[width=1\textwidth]{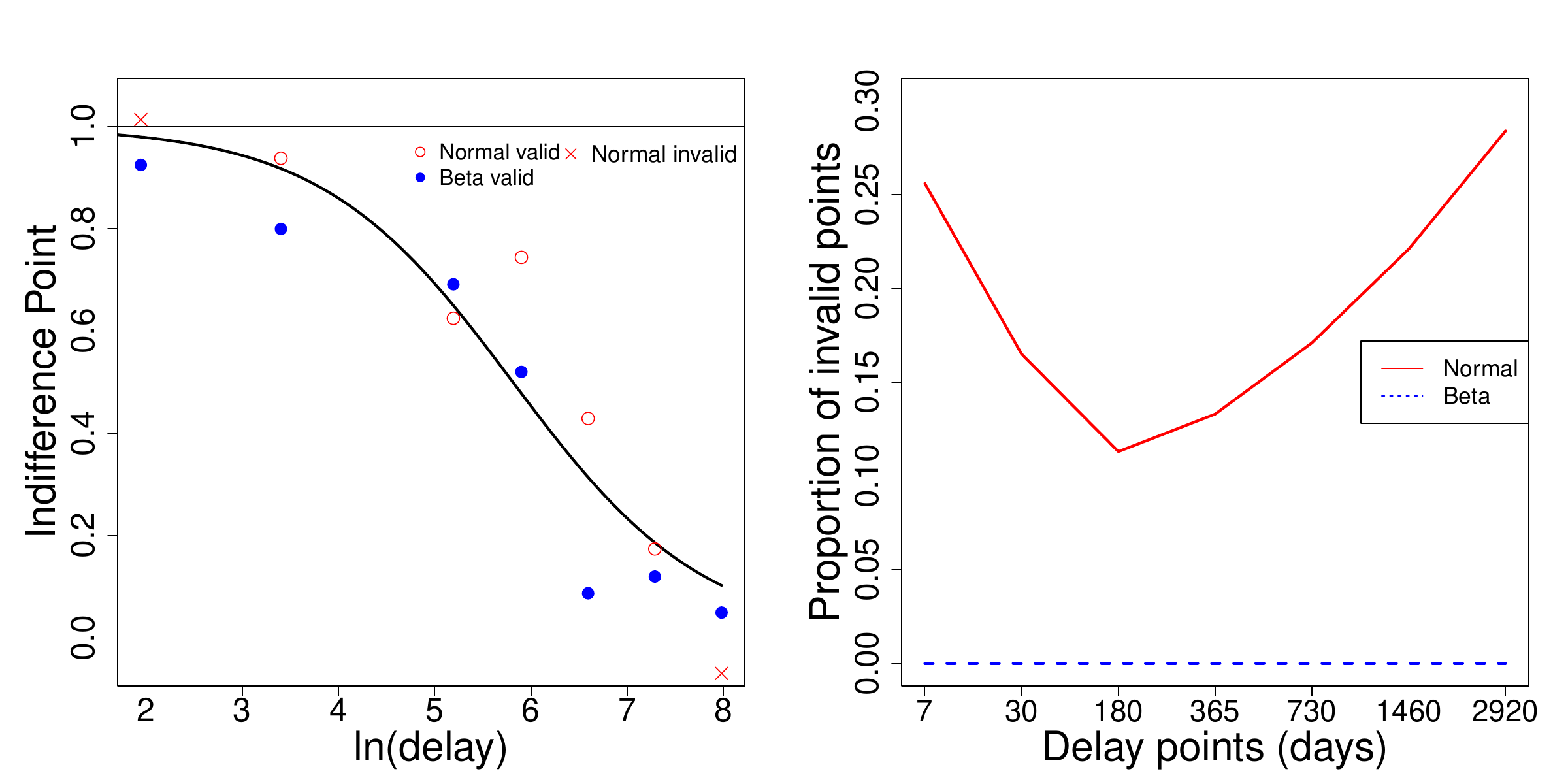}
    \caption{Left panel is simulated data from normal distribution and beta distribution and right panel is proportion of invalid values at each delay by simulation type over all subject among 1000 replications.}
    \label{fig:fig7}
\end{figure}

We also simulated data from {the} normal distribution and beta distribution based on data from \cite{jarvis2019effects}. Figure \ref{fig:fig7} {shows an example data set that shows that simulation} {from the} normal distribution gives invalid values, while {the} beta distribution gives valid data. The right panel of Figure \ref{fig:fig8}  shows the Monte Carlo simulation result. This shows the proportion of invalid values at each delay for all subjects among 1000 repetitions when data {are} simulated from {the} normal distribution.  Within the simulation study, there are 126000 simulated participants and 126000*7 simulated indifference points. The proportion of subjects having at least one invalid simulated indifference points is 0.768 and the proportion of invalid simulated indifference points is 0.1918.

%% file: Discussion.tex
\section{Discussion}

{We have proposed a nonlinear regression model that utilizes the beta distribution in order to better account for non-constant variance in indifference points and obey the [0,1] boundaries for such data. We include an {scale-location-truncation} modification to the beta PDF that allows the distribution to accommodate observed values of 0 and 1 while being virtually indistinguishable from the usual beta distribution everywhere else. We have compared this approach with  NLS to clearly show  the limitations of NLS and advantages of using SLT beta regression.} {Figures \ref{fig:fig4} and  \ref{fig:fig5} show that NLS-based methods for estimating $k$ are reliable and there is no need to reanalyze all previous discounting data when estimating $k$ is the only goal.} {The agreement between estimates of $ln(k)$ between NLS and SLT beta regression indicate that our methodology does not call previous analyses of discounting rate into question; rather we hope we have added the ability to more carefully characterize variability and simulate more realistic data in future studies.}

{We note that the previous analyses does not consider} standard beta regression, {which} has its limitation that it cannot estimate $k$ values when there exist a subject which has 0 or 1 indifference point. {For those who are interested, we include a comparison of ``vanilla'' beta regression and SLT beta regression in the Appendix. Of course, this analysis only includes the subset of particiopants who do not exhibit 0s or 1s in their data.}  As shown in the Figure \ref{fig:fig8} in Appendix, the PDF of beta distribution and SLT beta distribution looks identical to the naked eye. Therefore, SLT beta regression has advantages of beta regression and does not have limitation that beta regression has; it can estimate $k$ values when the subjects has 0 or 1 values for indifference points. 

We also illustrated simulated data {in} Figure \ref{fig:fig7}. This an example underscores the superiority of using beta model over standard normal approach. Since beta distribution is bounded between 0 and 1 (including 0 and 1), simulation from beta model always generate more human-like data. When we refer to human-like data, we imply that it falls within the range from 0 to 1 [0, 1]. This is because the questionnaire was intentionally crafted to prevent values exceeding 1 or going below 0. This method shows similar performance with NLS regression and has advantage in terms of simulations. 

{The proposed SLT beta regression approach does have some limitations and opportunities for future research. The beta regression models described in this paper target the analysis of delay discounting data that can be sensibly normalized to the [0,1] interval. Tasks that have a fixed upper amount meet this criterion. However, tasks that have indifference points on the [0, $\infty$) range, such as those arising from certain cross-commodity discounting procedures, would not seem to fit sensibly into the methods we propose. Another opportunity for future research is to use the ability of SLT beta regression to quantify non-constant variance to further study patterns of variability in delay discounting data. Finally, an additional future research goal is to fully deploy SLT beta regression in a hierarchical model that simultaneously estimates fixed and random effects.}


%% file: Appendix.tex
\newpage

\section*{Appendix}

\subsection*{Beta distribution}

The reparameterization proposed by \cite{ferrari2004beta} expresses the beta density in terms of a mean parameter {$\mu$} and scale parameter {$\phi$}.  With $\mu=\alpha/(\alpha+\beta)$ and $\phi=\alpha+\beta$:
\begin{equation}
     f(y;\mu, \phi)= \frac{\Gamma(\phi)}{\Gamma(\mu\phi)\Gamma((1-\mu)\phi)}y^{\mu\phi-1}(1-y)^{(1-\mu)\phi-1}, \ \  0\leq y \leq 1
\end{equation}\label{eq:eq4}
{where} $0 < \mu < 1$ and $\phi > 0$. \cite{cribari2010beta} write y $\sim$ $B(\mu, \phi)$ where $\text{E}(y)=\mu$ and $\text{VAR}(y)= \mu(1-\mu)/(1+\phi)$.

{The likelihood function associates the observed data with unknown parameters in the context of a probability distribution}. The log-likelihood is obtained by taking the natural logarithm of the likelihood function. 

The log-likelihood function {for a single data point $y$} is 
\begin{equation*}
   L(y;\mu, \phi)=ln(\Gamma(\phi))-ln(\Gamma(\mu\phi))-ln(\Gamma((1-\mu)\phi))
+(\mu\phi-1)ln(y)+((1-\mu)\phi-1)ln(1-y).
\end{equation*}

The log-likelihood function can not accommodate the 0 and 1 value because $ln(0)$ is undefined.

\subsection*{Scale location truncated beta regression details}

\begin{figure}[H]
  \centering
  \begin{subfigure}[b]{0.45\textwidth}
    \centering
    \includegraphics[width=\textwidth]{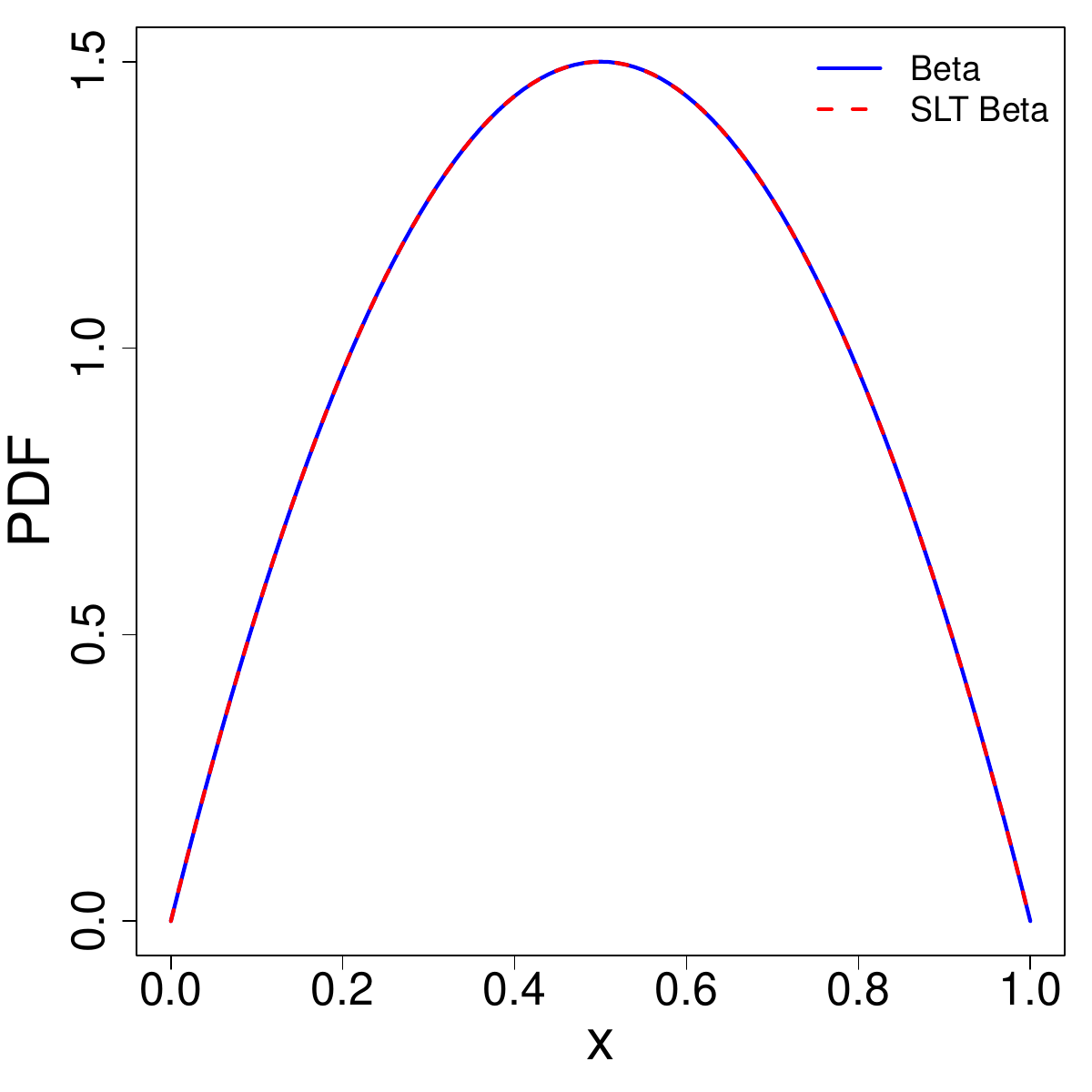}
  \end{subfigure}
  \hfill
  \begin{subfigure}[b]{0.45\textwidth}
    \centering
    \includegraphics[width=\textwidth]{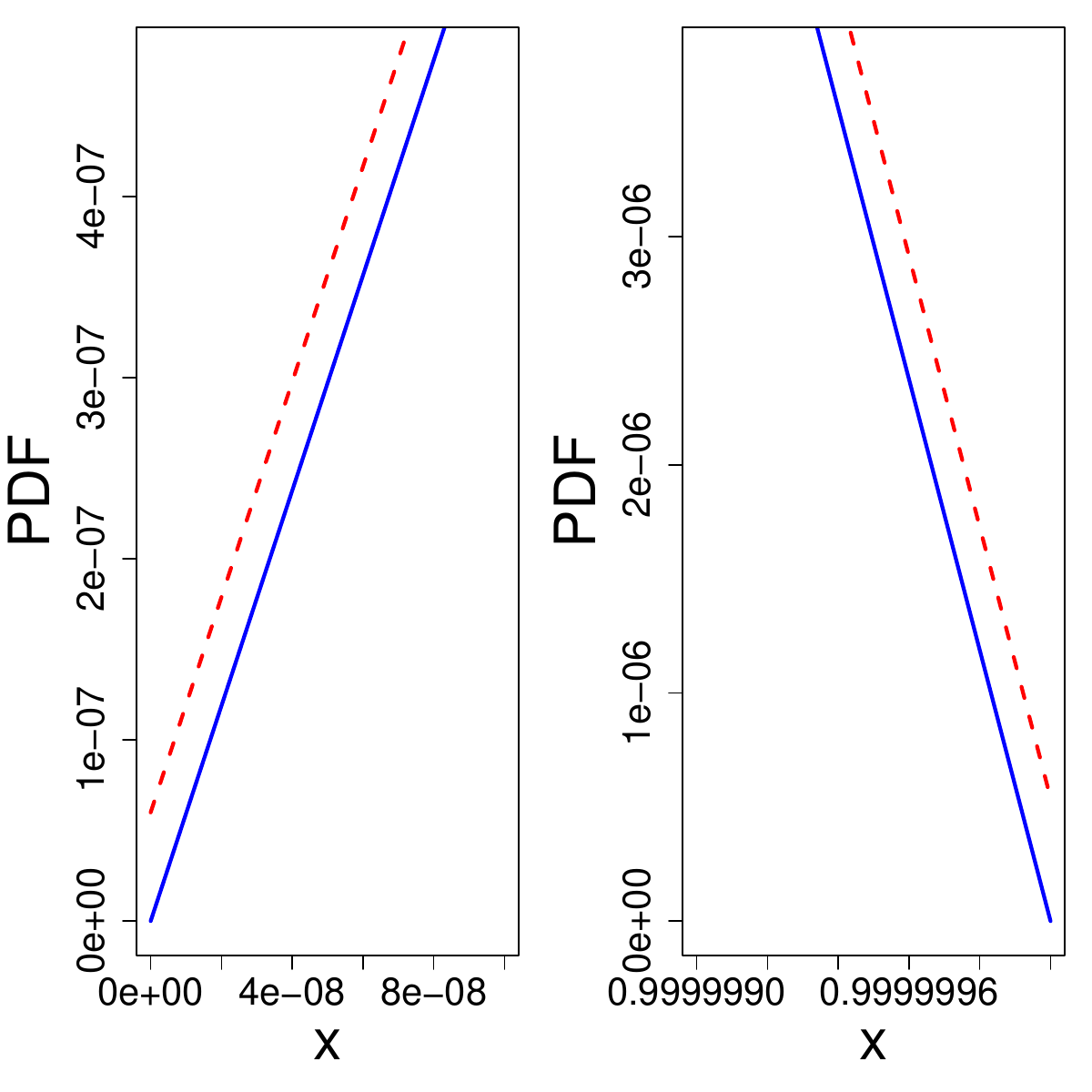}
  \end{subfigure}
  \caption{Left panel is the probability density function (PDF) of SLT beta and beta distribution. We can two density functions looks very close to each other. The right panel shows the region near 0 and 1. While PDF of beta is 0 in when x-axis is 0 and 1, SLT beta distribution has bigger value of PDF than 0.} 
  \label{fig:fig8}
\end{figure}

{By using a scale parameter $l$ and a location parameter $s$}, the beta density can be {modified into the SLT beta density, which can accommodate values of 0 and 1.} We chose scale parameter $s$ and location parameter $l$ that make SLT beta distribution and beta distribution be very close as seen in Figure \ref{fig:fig8}. {Let $z=(y-l)\cdot s$, where $z$ is location scale transformed form of $y$, where $y$ follows a beta distribution}. {The PDF of $z$ is:}

\begin{equation}
    f_Z(z;\alpha, \beta, s ,l)=\frac{\Gamma(\alpha+\beta)}
{\Gamma(\alpha)\Gamma(\beta)}\left(\frac{z}{s}+l\right)^{\alpha-1}\left(1-\left(\frac{z}{s}+l\right)\right)^{\beta-1}\frac{1}{s}
\label{eq:eq5}
\end{equation}

where $z$ = $(y-l)\cdot s$.  $s$ is scale parameter and $l$ is location parameter. 

The range of $z$ is [-$ls$, (1-$l$)$s$], which is not consistent {with} the [0,1] range of indifference points. To restrict the range of z to [0, 1], we normalize Equation (\ref{eq:eq5}) by dividing it with normalizing constant $F_{z}(1)-F_{z}(0)$. {$F_{z}(t)$ is a cumulative density function (CDF), which is the area under the probability density function (PDF) from the lower bound to the point $t$.}

{
\begin{equation*}
    F_z(t)=  P(Z \leq t) = \int_{-s\cdot l} ^{t} f_z(t)dt =
    \int_{-s\cdot l} ^{t} \frac{1}{B(\alpha,\beta)}\left(\frac{z}{s}+l\right)^{\alpha-1}\left( 1-\left(\frac{z}{s}+l\right)\right)^{\beta-1}\frac{1}{s} \ dz  
\end{equation*}
\begin{equation}
  =\int_{0} ^{\frac{t}{s}+l} \frac{1}{B(\alpha,\beta)} y ^{\alpha-1} (1-y)^{\beta-1}dy = P(Y  \leq  \frac{t}{s} + l)  =F_y \left({\frac{t}{s}+l}\right).
\end{equation}
}

Dividing the PDF of $z$ by $F_{z}(1)-F_{z}(0)$ makes PDF integrate to 1 over the interval [0, 1] when we restrict the range as [0,1]. Equation (\ref{eq:eq8}) is final form of scale location truncated beta density. 

\begin{equation}
    f_G(g)= \frac{f_Z(g)}{F_z(1)-F_z(0)}=\frac{\frac{\Gamma(\alpha+\beta)}
{\Gamma(\alpha)\Gamma(\beta)}\left(\frac{g}{s}+l\right)^{\alpha-1}\left(1-\left(\frac{g}{s}+l\right)\right)^{\beta-1}\frac{1}{s}}{F_y(\frac{1}{s}+l)-F_y(l)} ,\ \ 0\leq g \leq 1.
\label{eq:eq6}
\end{equation}

Using SLT beta regression, the response variable range expands to [0,1]. Since the SLT beta distribution has density mass at 0 and 1, we can optimize the log-likelihood function to estimate parameters.

By using the reparameterization mentioned in Section 2.1, we will have Scale Location Truncated beta density in Equation (\ref{eq:eq8}).

\begin{equation}
    f_G(g)= \frac{\frac{\Gamma(\phi)}{\Gamma(\mu\phi)\Gamma((1-\mu)\phi)}(\frac{g}{s}+l)^{\mu\phi-1}(1-(\frac{g}{s}+l))^{(1-\mu)\phi-1}}{F_y(\frac{1}{s}+l)-F_y(l)}, \ \ 0 \leq g \leq 1.
    \label{eq:eq8}
\end{equation}

Equation (\ref{eq:eq8}) is location-scale beta distribution density which SLT is based on. {In Equation (\ref{eq:eq8}), we introduce a reparameterization of the response variable, denoted as $g$. In the main document, we use $y$ as the outcome variable.}

\subsection*{Standard beta regression}

\begin{figure}[H]
    \centering
    \includegraphics[width=1\textwidth]{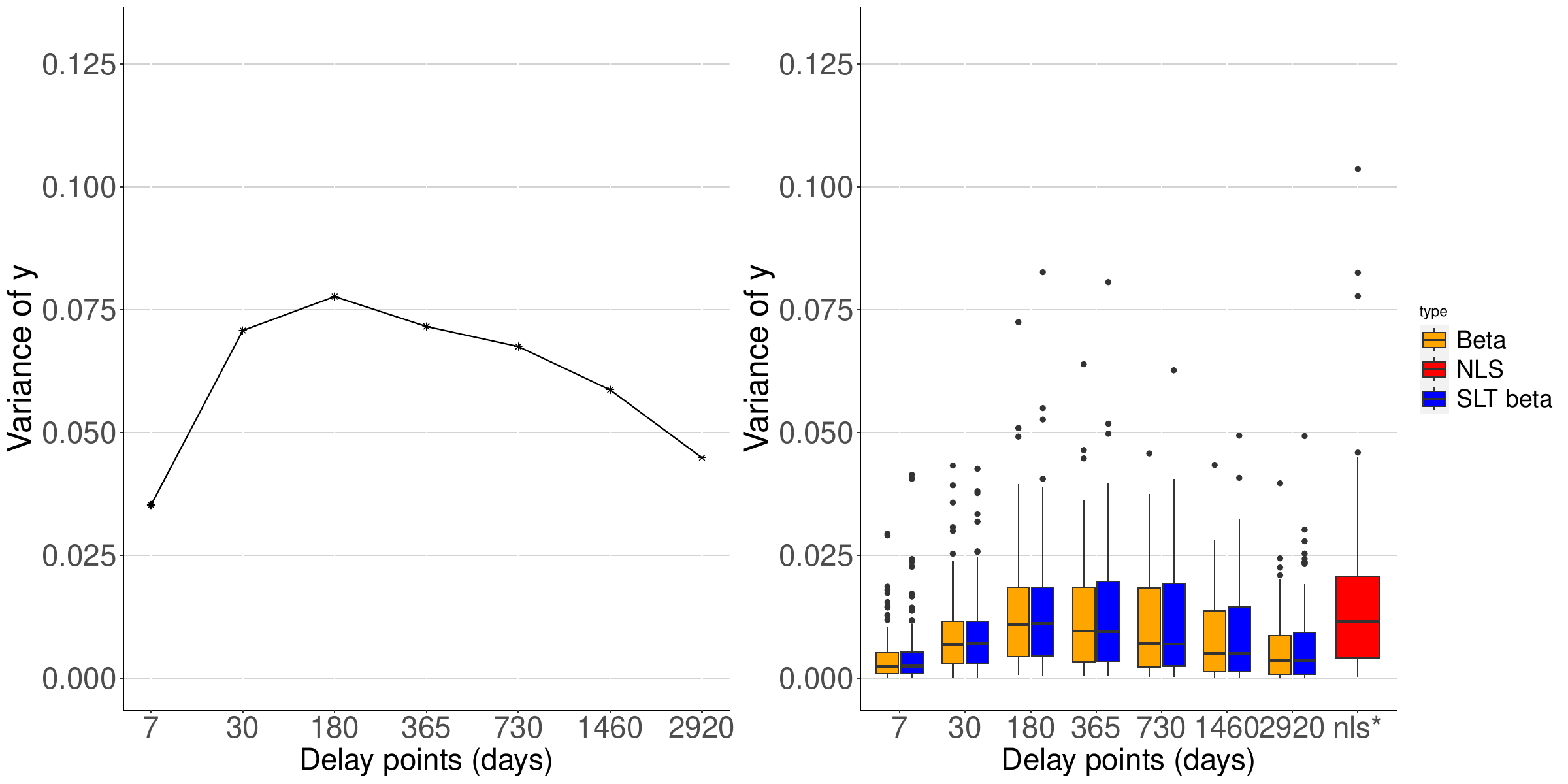}
    \caption{Variance among subjects at each delay (left). Box plot of model estimated variance from SLT beta regression and standard beta regression at each delay and variance from NLS (right). Delays are graphed as equally spaced for ease of visualization.}
    \label{fig:fig9}
\end{figure}

\begin{figure}[H]
    \centering
    \includegraphics[width=1\textwidth]{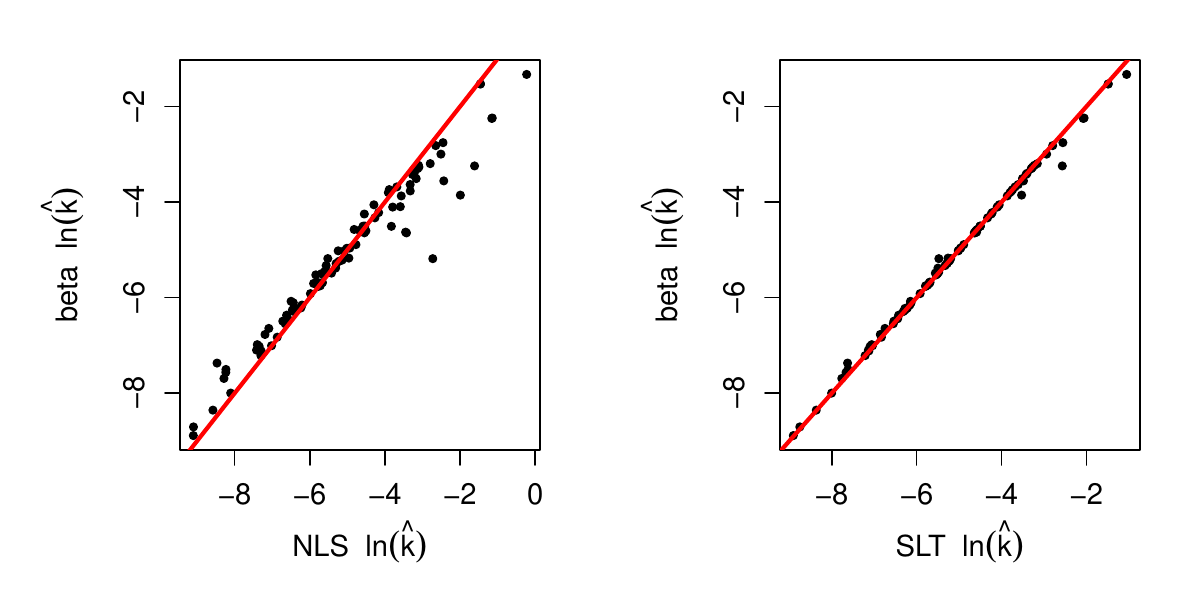}
    \caption{Scatter plots of ln($\hat{k}$). In left panel, horizontal axis is ln($\hat{k}$) from NLS and vertical axis is ln($\hat{k}$) from original beta regression. In right panel, horizontal axis is ln($\hat{k}$) from SLT beta regression and vertical axis is ln($\hat{k}$) from original beta regression. Red line indicates y=x.}
    \label{fig:fig10}
\end{figure}

{For the reader who is curious about the standard (i.e., non-SLT) beta regression, Figures \ref{fig:fig9} and \ref{fig:fig10} show results similar to those presented in Section 3. Since standard beta regression cannot accommodate 0 and 1 indifference point values, we selected participants who pass Johnson-Bickel Criteria and whose indifference points do not include 0 or 1.}

{In Figure \ref{fig:fig9}, the model estimated variance from beta and SLT beta regression are very similar. Figure \ref{fig:fig9} is similarto Figure \ref{fig:fig4} but the estimated variance from beta regression is added. In addition, to make the estimated $k$ value comparison clearer, we have added the Figure \ref{fig:fig10}. In the left panel of the plot, it similar to Figure \ref{fig:fig5}  but instead of comparing the estimated $ln(\hat{k})$ of SLT beta and NLS, we compared that of beta and NLS. In the right panel of Figure \ref{fig:fig10} , We compared the estimated  $ln(\hat{k})$ of SLT beta and beta regression, and they are remarkably close.}

R code used for this paper can be found in \href{https://bitbucket.org/paper-code-kmg/slt_beta/src/main/}{here} at \url{https://bitbucket.org/paper-code-kmg/slt_beta/src/main/}